\documentclass[aps,prl,preprint,showpacs,amsmath,floatfix,superscriptaddress]{revtex4}

\usepackage{graphicx}

\begin{document}

\title{Coulomb-hole summations and energies for GW calculations with limited number of empty orbitals: a modified static remainder approach}

\author{Jack Deslippe}
\affiliation{Department of Physics, University of California at
Berkeley, California 94720 and Materials Sciences
Division, Lawrence Berkeley National Laboratory, Berkeley,
California 94720}

\author{Georgy Samsonidze}
\affiliation{Department of Physics, University of California at
Berkeley, California 94720 and Materials Sciences
Division, Lawrence Berkeley National Laboratory, Berkeley,
California 94720}

\author{Manish Jain}
\affiliation{Department of Physics, University of California at
Berkeley, California 94720 and Materials Sciences
Division, Lawrence Berkeley National Laboratory, Berkeley,
California 94720}

\author{Marvin L. Cohen}
\affiliation{Department of Physics, University of California at
Berkeley, California 94720 and Materials Sciences
Division, Lawrence Berkeley National Laboratory, Berkeley,
California 94720}

\author{Steven G. Louie}
\affiliation{Department of Physics, University of California at
Berkeley, California 94720 and Materials Sciences Division, Lawrence Berkeley
National Laboratory, Berkeley, California 94720}

\date{\today}

\begin{abstract}
{\em Ab initio} GW calculations are a standard method for computing the spectroscopic properties of many materials. The most computationally expensive part in conventional implementations of the method is the generation and summation over the large number of empty orbitals required to converge the electron self energy. We propose a scheme to reduce the summation over empty states by the use of a modified static-remainder approximation, which is simple to implement and yields accurate self energies for both bulk and molecular systems requiring a small fraction of the typical number of empty orbitals.
\end{abstract}


\maketitle


\section{Introduction}

The GW methodology \cite{hedin65,hedin70,hybertsen86} has been successfully applied to the study of the quasiparticle properties of a wide range of systems \cite{louiebook06} from traditional bulk semiconductors, insulators and metals to nanosystems like polymers, nanotubes and molecules \cite{spataru04,spataru04long,deslippe07}. The approach yields quantitatively accurate quasiparticle band gaps and dispersion relations from first-principles. A perceived drawback of the GW methodology is its computational cost; usually thought to be an order of magnitude more than a typical DFT calculation. One of the main computational bottlenecks of the traditional {\it ab intio} GW method \cite{hybertsen86} is the cost to generate the large number of empty orbitals needed to converge the Coulomb-hole summation term of the self-energy.  

Within the conventional GW approach, the quasiparticle energies and wavefunctions (\textit{i.e.}, the one-particle excitations) are computed by solving the following Dyson equation \cite{hedin70,hybertsen86} (in atomic units):
\begin{equation}
\label{dyson_equation}
\left[ -\frac{1}{2}\nabla^2+V_{\rm ion}+V_{\rm H}+{\it\Sigma}(E_{n\bf k}^{\rm QP}) \right] \psi_{n\bf k}^{\text{QP}}=E_{n\bf k}^{\rm QP}\psi_{n\bf k}^{\text{QP}}
\end{equation}
where $\Sigma$ is the non-local, energy-dependent, self-energy operator within the GW approximation, and $E_{n\bf k}^{\rm QP}$ and $\psi_{n\bf k}^{\rm QP}$ are the quasiparticle energies and wavefunctions, respectively. In the typical GW approach, density functional theory (DFT) within the Kohn-Sham approximation \cite{kohn65} is often chosen as the starting point for a subsequent calculation of the electron self-energy: the Kohn-Sham \cite{kohn65} wavefunctions and eigenvalues are used here as a first guess for their quasiparticle counterparts. In principle, Eq. \ref{dyson_equation} is a matrix equation, where $\Sigma$ should be constructred in an appropriate basis. In many cases, only the diagonal elements are sizable within the basis spanned by the Kohn-Sham mean-field orbitals. We assume this to be the case for the rest of the paper. The effects of $\Sigma$ can thus be treated within first-order perturbation theory in the form $\Sigma = V_{\rm xc} + (\Sigma - V_{\rm xc})$, where $V_{\rm xc}$ is the independent-particle mean-field approximation to the exchange-correlation potential of Kohn-Sham system \cite{kohn65}.

Within the GW approximation, the self-energy operator is expressed as $\Sigma=iGW$, where $G$ is the electronic Green's function and $W$ is the dynamically screened Coulomb interaction. In the GW and static-COHSEX (the static limit of GW) approximations for the self energy, the self-energy operator, $\Sigma$, can be broken into two parts, \cite{hybertsen86,hedin70} $\Sigma = \Sigma_{\rm SX} + \Sigma_{\rm CH}$ where $\Sigma_{\rm SX}$ is the screened-exchange operator and $\Sigma_{\rm CH}$ is the Coulomb-hole operator. The screened-exchange operator is similar to the Fock operator in Hartree-Fock theory, except the bare Coulomb interaction is replaced by the screened Coulomb interaction: $W_{{\bf GG}'}{\left({\bf q}\;\!;\omega\right)}=\epsilon^{-1}_{{\bf GG}'}{\left({\bf q}\;\!;\omega\right)}v({\bf q}+{\bf G}')$ where $v$ is the bare Coulomb interaction. When $G$ and $W$ are constructed in a non-self-consistent way from the DFT orbitals and eigenvalues, the $GW$ approach is referred to as being within the $G_0W_0$ approximation. In this article, all results are presented within this approximation.

In a conventional GW calculation within the generalized plasmon-pole approximation \cite{hybertsen86}, both the calculation of the Coulomb-hole self energy term:
\begin{widetext}
\begin{eqnarray}
\label{ch-sum}
\left<n{\bf k}\right|\Sigma_{\rm CH}^N{\left({\bf r},{\bf r}';E\right)}\left|n{\bf k}\right>=
\frac{1}{2}\sum_{n''}^N\sum_{{\bf qGG}'}
\left<n{\bf k}\right|e^{i({\bf q}+{\bf G})\cdot{\bf r}}\left|n''{\bf k}{-}{\bf q}\right>
\left<n''{\bf k}{-}{\bf q}\right|e^{-i({\bf q}+{\bf G}')\cdot{\bf r}'}\left|n{\bf k}\right>
\\ \nonumber \times\,\{ \frac{\Omega^2_{{\bf GG}'}{\left({\bf q}\;\!\right)}}
{\tilde{\omega}_{{\bf GG}'}{\left({\bf q}\;\!\right)}
\left[E\,{-}\,E_{n''{\bf k}{-}{\bf q}}{-}\,
\tilde{\omega}_{{\bf GG}'}{\left({\bf q}\;\!\right)}\right]}
\;v{\left({\bf q}{+}{\bf G}'\right)} \}
\end{eqnarray}
\end{widetext}
and the calculation of the dielectric screening matrix, $\epsilon = 1 + 4 \pi \chi$, at $\omega = 0$:
\begin{align}
\label{chi}
\epsilon_{{\bf GG}'}&{\left({\bf q}\;\!;0\right)} =
\delta_{{\bf GG}'} \\
\nonumber
&\,{-}\,v{\left({\bf q}{+}{\bf G}\right)}
\,\,{}\sum_{n}^{\rm occ}\sum_{n'}^{N}\sum_{{\bf k}} 
\left<n{\bf k}{+}{\bf q}\right|e^{i({\bf q}+{\bf G})\cdot{\bf r}}\left|n'{\bf k}\right> \\
\nonumber
&\left<n'{\bf k}\right|e^{-i({\bf q}+{\bf G}')\cdot{\bf r}'}\left|n{\bf k}{+}{\bf q}\right> \times\frac{1}{E_{n{\bf k}{+}{\bf q}}\,{-}\,E_{n'{\bf k}}},
\end{align}
involve a summation over empty orbitals.  Here $N$ is the number of empty orbitals in the truncated sum, $n\bf{k}$ is a Bloch orbital with a given crystal momentum $\bf{k}$, band index $n$ and energy $E_{n\bf{k}}$, $v(\bf{q}+\bf{G})$ is the bare Coulomb interaction in reciprocal space 
and $\Omega_{\bf{GG}'}(\bf{q})$ and $\tilde{\omega}_{\bf{GG}'}(\bf{q})$ are plasmon-pole parameters \cite{hybertsen86}. 	

There has been much research effort invested in recent years to reduce the need for empty orbitals in the GW formalism \cite{reining97, umari09, giustino10, bruneval08, tiago06, gsm10CAR, berger10,steinbeck00,sib-pwsub}. The approaches by Umari et. al. \cite{umari09} and Giustino et. al. \cite{giustino10} eliminate the need for empty states entirely by constructing the dielectric response and self-energy from only occupied states within a linear-response Sternheimer equation appraoch \cite{giustino10}. While these approaches eliminate the need for empty orbitals, they are conceptually more complicated as well as more complicated to implement and optimize. It is therefore still of great value to address the computational cost of the empty orbital generation within the traditional GW approach laid out above. 

One approach to addressing the problem of the cost associated with empty orbitals in the traditional GW approach is to approximate the true DFT empty orbitals with approximate orbitals that are computationally cheaper to generate \cite{gsm10CAR,steinbeck00,sib-pwsub}. In the recent work of Samsonidze et. al. \cite{gsm10CAR}, the authors proposed replacing the expensive step of constructing the exact Kohn-Sham empty states from a traditional DFT package with a computationally inexpensive process of constructing the empty states from a reduced basis set consisting of plane-waves and resonant orbitals (generated in SIESTA \cite{soler02siesta}) orthogonalized to the real occupied Kohn-Sham orbitals. While it was shown that this approach vastly reduces the cost of generating the required empty orbitals, the approach adds significantly to the complication of the GW process. In particular, one now must run a traditional plane wave DFT calculation, a local orbital DFT calculation and a post-processing orthogonalization step in order to generate the required electron orbitals needed to proceed to the GW calculation. Additionally, the explicit sums in Eq. \ref{ch-sum} and \ref{chi} must still be performed over these orbitals.

Another approach to the empty state problem was proposed by Tiago et. al. \cite{tiago06}: a truncation of the sum over empty orbitals in Eq. \ref{ch-sum} can be achieved with minimal loss of accuracy by adding the contribution of the remaining orbitals within the static (COHSEX) approximation \cite{hedin70,hybertsen86}.  The idea relies on the fact that, in static-COHSEX, unlike GW, the Coulomb-hole energy can be written in a simple closed form (see next section) as well as in a sum over empty-states, as the static limit of Eq. \ref{ch-sum}. It was proposed that one may approximate the missing Coulomb-hole contribution when truncating the sum after $N$ (i.e. the contribution to the sum from the empty orbitals with index between $N$ and $\infty$) in a GW calculation by their contribution to the Coulomb-hole energy in a static-COHSEX calculation.

This static approximation, however, was shown to be of limited use by Bruneval et. al. \cite{bruneval08}, where, instead of using the static approximation for the remaining part of the Coulomb-hole sum, the authors proposed using an approach based on a common non-zero energy denominator in Eq. \ref{ch-sum}. If a constant denominator is assumed, than one may use the completion relation:
\begin{equation}
\sum_{n=N+1}^{\infty} |n {\bf k} \rangle \langle n {\bf k} | = 1 - \sum_{n=1}^{N} |n {\bf k} \rangle \langle n {\bf k} |
\end{equation}
to replace the sum over the missing empty orbitals with a sum over the available orbitals. In order to apply this directly to Eq. \ref{ch-sum} and Eq. \ref{chi}, one must replace the $n$ dependent denominator with a constant. The main drawback of this common energy denominator approximation (CEDA) approach (known also as the extrapolar method) is that the energy denominator is not uniquely defined and can only be treated as a somewhat ad-hoc parameter, and the quasiparticle energy convergence is not monotonic with this parameter.  

Recent studies by Kang and Hybertsen \cite{kang10} have shown that a modified static COHSEX approach can be used to accurately minimize the empty orbitals problem in the Coulomb-hole summation of Eq \ref{ch-sum}. In that work, the authors propose completely replacing the GW Coulomb-hole operator with a closed-form static operator (similar to that used by Tiago \cite{tiago06}) with a ${\bf q}$ dependent coefficient, $f({\bf q})$, fit to match the GW result. This approach has the advantage of being completely closed form, but can be improved if one relaxes the requirement that the modified operator is used to replace the true GW contributions not only from high energy empty orbitals but also from valence and low energy conduction orbitals.

In this article, we propose a modified static remainder approach based on Tiago's results \cite{tiago06} that is more fully justified by the recent Kang-Hybertsen result \cite{kang10}. The new approach yields accurate GW Coulomb-hole absolute energies (to within 100 meV) with less than 10\% of the traditionally necessary empty orbitals. Furthermore, unlike the extrapolar method of Bruneval \cite{bruneval08}, this approach yields an easy to implement procedure with no adjustable parameters. For simplicity of presentation, we shall discuss our approach within the generalized plasmon pole model for the dielectric matrix. This approach can be straightfowardly applied to any existing GW computational package.

\section{Traditional GW Convergence with Empty Orbitals}

\begin{figure}
\includegraphics[width=6.324cm] {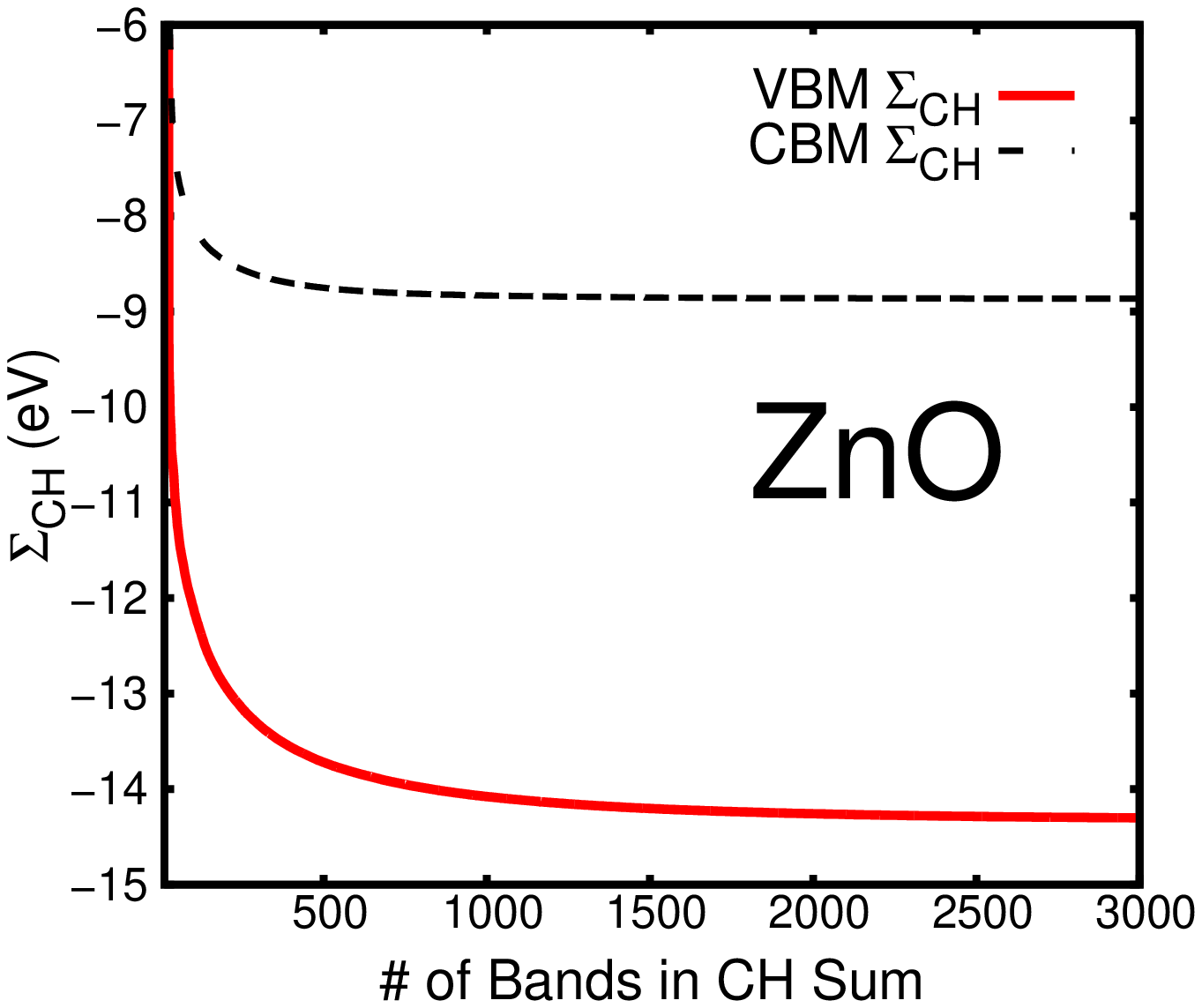} \\
\includegraphics[width=6.324cm] {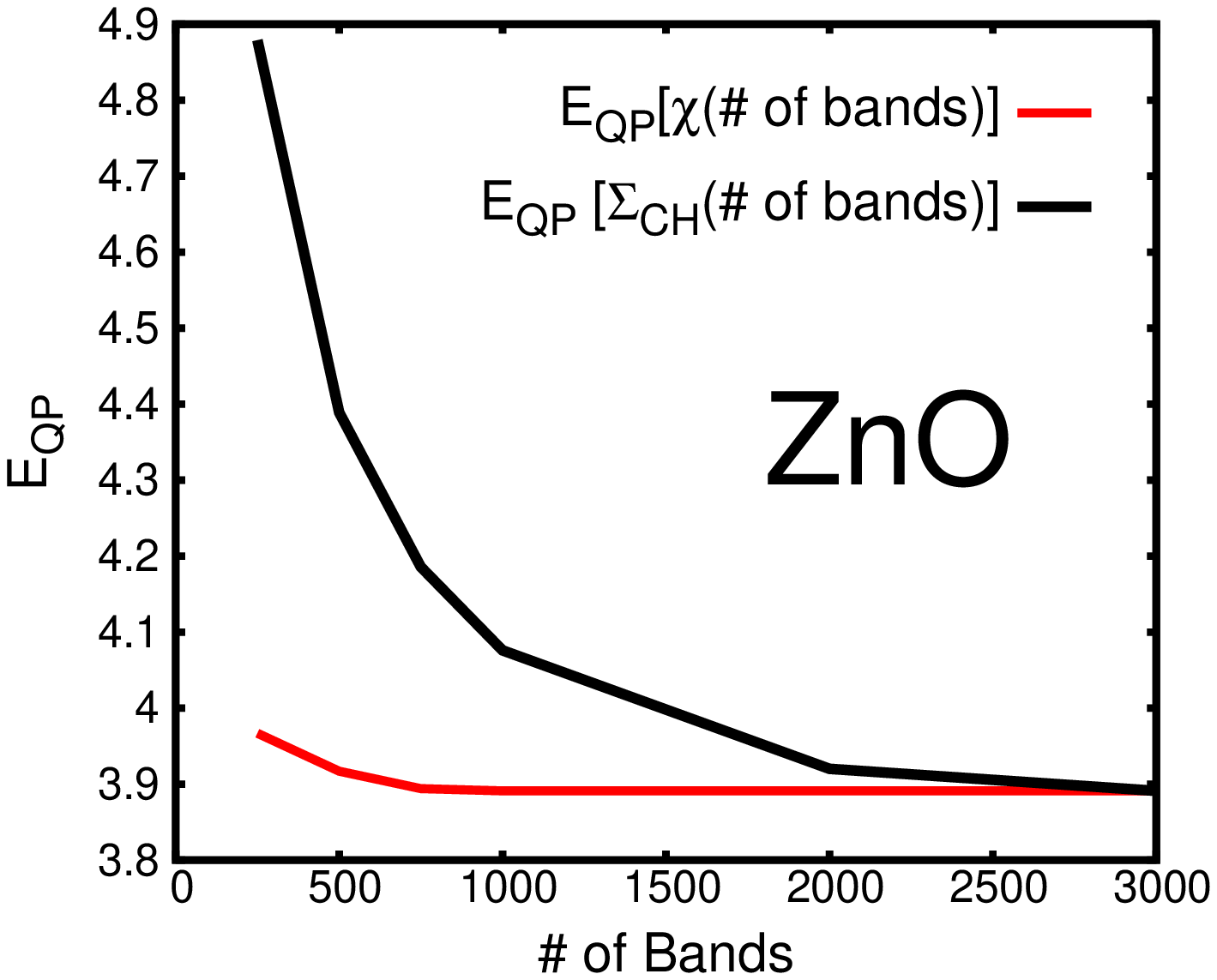}
\caption{Top: The convergence of the Coulomb-hole contribution to the self-energy, Eq. \ref{ch-sum}, with respect to the number of orbitals included in the summation, $N$, using a dielectric matrix calculated with 1000 empty bands. For all calculations on ZnO, a 5x5x4 {\bf k}-point grid is used. Bottom: The convergence of the quasiparticle energy, $\text{E}_{\text{QP}}$, with respect to empty states in the polarizability sum Eq. \ref{chi} and with respect to empty states in the Coulomb-hole sum Eq. \ref{ch-sum}. The red curve shows the VBM $\text{E}_{\text{QP}}$ in ZnO using a fixed 3,000 bands in the Coulomb-hole summation and varying the number of bands included in the polarizability summation. The black curve shows the VBM $\text{E}_{\text{QP}}$ in ZnO using a fixed 1,000 bands in the polarizability summation and varying the number of bands included in the Coulomb-hole summation.
}
\label{fig5}
\end{figure}

As mentioned in the introduction, the band convergence of absolute energy levels in $\Sigma_{\text{CH}}$ with the number of empty orbitals is extremely slow. In principle, one must converge a GW calculation with respect to the number of empty stats in both Eq. \ref{chi} and \ref{ch-sum}. However, for many systems, the quasiparticle energy dependence on the number of empty orbitals in the dielectric screening (e.g. Eq. \ref{chi}) converges much faster than with the number of empty orbitals in Eq. \ref{ch-sum}. For example, recent calculations for ZnO show that the Coulomb-hole contribution to the electronic band gap does not converge until 3,000+ empty orbitals are included in the summation in Eq. \ref{ch-sum} \cite{shih10}. In Fig. \ref{fig5}, we demonstrate the slow convergence of Eq. \ref{ch-sum} in ZnO. The situation is even worse for nanosystems where absolute energies are often required for applications involving interfaces over which absolute energy level alignment is needed such as the cases for molecular electronics or photovoltaic applications.  

From the bottom panel of Fig. \ref{fig5}, it is immediately evident that the quasiparticle energy converges much more slowly with respect to the number of empty orbitals included in the Coulomb-hole summation, Eq. \ref{ch-sum} than from the $\epsilon$ summation, Eq. \ref{chi}. Additionally, one may compute Eq. \ref{chi} in an alternative density functional perturbation theory approach that avoids the sum over empty orbitals.  Similar techniques \cite{giustino10,umari10} to avoid the empty orbitals for Eq. \ref{ch-sum} have been proposed but they are more difficult to implement and use. Therefore, any reduction in the summation in Eq. \ref{ch-sum} over large numbers of empty orbitals can greatly reduce the cost of calculation for standard GW approaches.

\section{Methodology}

The static COHSEX method is the static limit of the GW approximation for the self energy -- where everywhere $\epsilon(\bf{G},\bf{G}',\omega)$ is replaced by $\epsilon(\bf{G},\bf{G}',0)$. The static remainder approach is based on the fact that the expectation value of the static COHSEX Coulomb-hole operator can be expressed either in a closed form or as a sum over empty orbitals:

\begin{widetext}
\begin{align}
\label{stat-sum}
& \Sigma_{\text{CH}}^{Coh/N}(n,{\bf k})= 
\\ \nonumber & \frac{1}{2}\sum_{n''}^N\sum_{{\bf qGG}'}
\left<n{\bf k}\right|e^{i({\bf q}+{\bf G})\cdot{\bf r}}\left|n''{\bf k}{-}{\bf q}\right>
\left<n''{\bf k}{-}{\bf q}\right|e^{-i({\bf q}+{\bf G}')\cdot{\bf r}'}\left|n{\bf k}\right> \nonumber \times  \{ \left[\epsilon_{{\bf GG}'}^{-1}{\left({\bf q}\;\!;0\right)}
\,{-}\,\delta_{{\bf GG}'}\right]
v{\left({\bf q}{+}{\bf G}'\right)} \}
\\ \nonumber
\end{align}
and,
\begin{align}
\Sigma_{\text{CH}}^{Coh/\infty}(n,{\bf k})=
\frac{1}{2}\sum_{{\bf qGG}'}
\left<n{\bf k}\right|e^{i({\bf G}-{\bf G}')\cdot{\bf r}}\left|n{\bf k}\right>
\left[\epsilon_{{\bf GG}'}^{-1}{\left({\bf q}\;\!;0\right)}
\,{-}\,\delta_{{\bf GG}'}\right]
v{\left({\bf q}{+}{\bf G}'\right)}
\label{stat-closed}
\end{align}
\end{widetext}
where $N$ and $\infty$ denote a truncated empty state summation and closed form expression, respectively.  Equation (\ref{stat-sum}) is equal to Eq. \ref{ch-sum} in the limit of static dielectric screening. 
In the work of Kang et. al. \cite{kang10}, the authors propose using a modified static-COHSEX operator that mimics the GW operator to entirely remove the need for empty orbitals. In our current approach, we include the full GW contribution from the low energy orbitals and add a single correction for the high-energy orbitals, where the static approximation is expected to perform well. One advantage of the present approach is that it can be used in conjunction with a full-frequency (as opposed to GPP model) screening approach to both calculate the fine structure of energy dependence of the self energy, $\Sigma(\omega)$, as well as converging the absolute value with respect to empty orbitals. In our modified static remainder approach, we calculate both the GW $\Sigma_{\text{CH}}$ partial sum (Eq. \ref{ch-sum}) and the COHSEX $\Sigma_{\text{CH}}$ partial sum (Eq. \ref{stat-sum}) up to the number of DFT bands available.  We then add a modified static correction to the GW Coulomb-hole energies:
\begin{widetext}
\begin{align}
\label{static-remainder}
\left<n{\bf k}\right| & \Sigma_{\rm CH}^\infty{\left({\bf r},{\bf r}';E\right)}\left|n'{\bf k}\right>= 
\\ \nonumber
& \left<n{\bf k}\right|\Sigma_{\rm CH}^N{\left({\bf r},{\bf r}';E\right)}\left|n'{\bf k}\right> +\frac{1}{2}
\bigg(\left<n{\bf k}\right|\Sigma_{\rm CH}^{Coh/\infty}{\left({\bf r},{\bf r}'\right)}\left|n'{\bf k}\right>-\left<n{\bf k}\right|\Sigma_{\rm CH}^{Coh/N}{\left({\bf r},{\bf r}'\right)}\left|n'{\bf k}\right>\bigg).
\end{align}
\end{widetext}
The factor of 1/2 in Eq. \ref{static-remainder} is justified from the recent work of Kang and Hybertsen \cite{kang10}, where the authors show that the GW contribution of high energy bands (corresponding to large $\bf{G}$-vectors) to the Coulomb-hole self energy asymptotes to 1/2 of the equivalent static COHSEX band contribution.  

One may qualitatively derive this result from Eq. \ref{ch-sum}, if one assumes that we are interested in a state $n$ with energy, $E$, near zero, and that, for a given high $n''$, the sum over matrix elements are dominated by a small set of ${\bf q + G}$ near such that $|\hbar({\bf q+G})|^2/2m \approx E_{n''}$ and that the plasma frequency for large ${\bf q}$ and ${\bf G}$ obeys a homogeneous gas dispersion $\tilde{\omega}_{{\bf G,G}}({\bf q}) \approx |\hbar({\bf q+G})|^2/2m$. In this case, the contribution to Eq. \ref{ch-sum} from a high-energy empty orbital $n''$ reduces to:
\begin{widetext}
\begin{eqnarray}
\left<n{\bf k}\right|\Sigma_{\rm CH}^N{\left({\bf r},{\bf r}';E\right)}\left|n'{\bf k}\right>=
\frac{1}{2}\sum_{{\bf qGG}'}
\left<n{\bf k}\right|e^{i({\bf q}+{\bf G})\cdot{\bf r}}\left|n''{\bf k}{-}{\bf q}\right>
\left<n''{\bf k}{-}{\bf q}\right|e^{-i({\bf q}+{\bf G}')\cdot{\bf r}'}\left|n'{\bf k}\right>
\\ \nonumber \times\,\{ \frac{\Omega^2_{{\bf GG}'}{\left({\bf q}\;\!\right)}}
{2 \tilde{\omega}^2_{{\bf GG}'}{\left({\bf q}\;\!\right)}
}
\;v{\left({\bf q}{+}{\bf G}'\right)} \}
\\ \nonumber
= \frac{1}{4}\sum_{{\bf qGG}'}
\left<n{\bf k}\right|e^{i({\bf q}+{\bf G})\cdot{\bf r}}\left|n''{\bf k}{-}{\bf q}\right>
\left<n''{\bf k}{-}{\bf q}\right|e^{-i({\bf q}+{\bf G}')\cdot{\bf r}'}\left|n{\bf k}\right> \nonumber 
\\ \nonumber \times  \{ \left[\epsilon_{{\bf GG}'}^{-1}{\left({\bf q}\;\!;0\right)}
\,{-}\,\delta_{{\bf GG}'}\right]
v{\left({\bf q}{+}{\bf G}'\right)} \},
\end{eqnarray}
\end{widetext}
which, when compared to Eq. \ref{stat-sum}, confirms the factor of $\frac{1}{2}$.  

The Coulomb-hole self-energy contribution to the convergence of energy levels in bulk silicon (using a 5x5x5 {\bf k}-point grid) and the silane (SiH$_{4}$) molecule (in a supercell calculation) are shown in Fig. \ref{fig2} as a function of the band cutoff, $N$, in Eq. \ref{ch-sum}.  The convergence on energy levels in silane is significantly slower than that in silicon \cite{rohlfing00} because of the large number of free-electron like vacuum states.  The silicon calculations were done with a 25 Rydberg wavefunction cutoff and a 10 Rydberg dielectric matrix cutoff.  The silane calculations were done with a 75 Rydberg wavefunction cutoff and a 6 Rydberg dielectric matrix cutoff.  The needed volume of the supercell used, $(25 au)^3$, and the corresponding number of vacuum states, is minimized by using a truncated Coulomb interaction \cite{beigi06}. 
Despite this, the largest computational cost in the GW calculation on silane is the DFT generation of the empty orbitals, representing more than 50\% of the total computational expense.  The calculation of the polarizability and the evaluation of the self energy require less computational time, and they scale nearly linearly to thousands of CPUs.

\section{Results}

\begin{figure*}
\includegraphics[width=6.324cm] {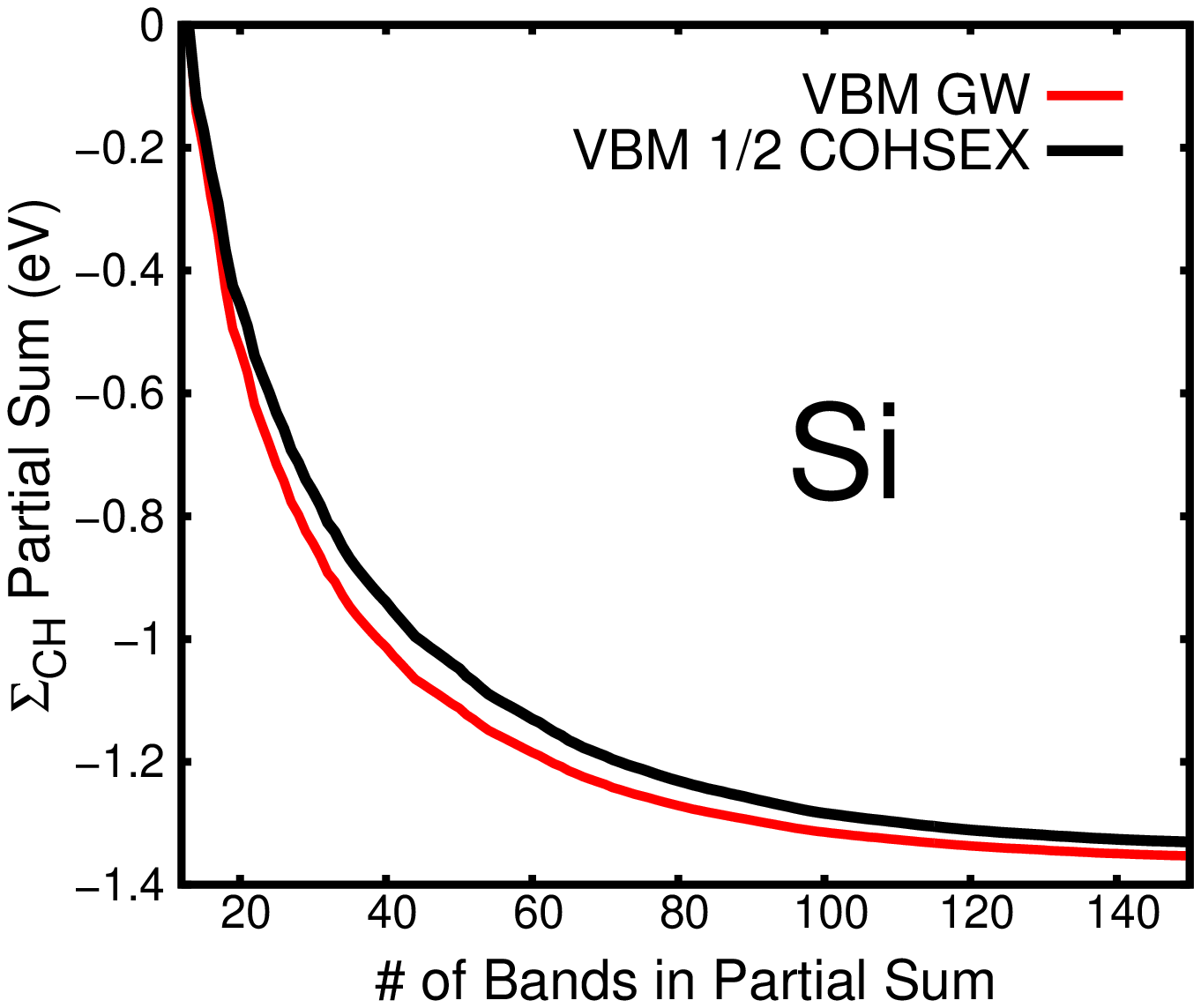}
\includegraphics[width=6.324cm] {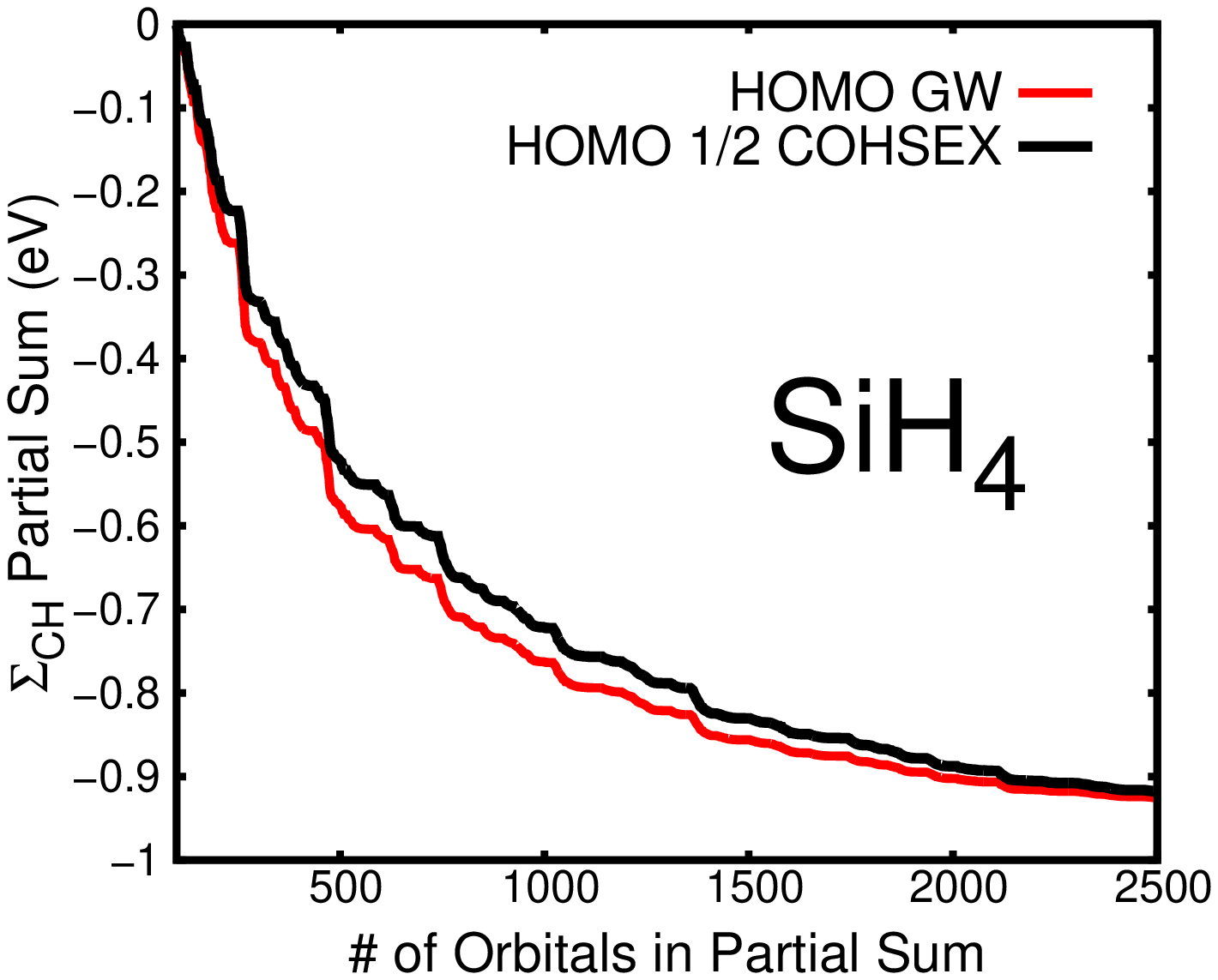}
\caption{Comparison between the contributions to the Coulomb-hole sum for the full GW operator vs.
results from the 1/2 the static COHSEX Coulomb-hole operator for orbitals beyond the number of real DFT bands/orbitals used: 12 in silicon and 100 in Silane. A 5x5x5 {\bf k}-point grid is used in Si. The plotted quantity is $\sum_{n''=n_{DFT}+1}^N\sum_{{\bf qGG}'}
\left<n{\bf k}\right|e^{i({\bf q}+{\bf G})\cdot{\bf r}}\left|n''{\bf k}{-}{\bf q}\right>
\left<n''{\bf k}{-}{\bf q}\right|e^{-i({\bf q}+{\bf G}')\cdot{\bf r}'}\left|n'{\bf k}\right>
\times \text{I}^{\text{CH}}_{{\bf GG}'}({\bf q},n,n',n'')$ where $\text{I}^{\text{CH}}$ is the term in $\{ \}$ in Eqs. (\ref{ch-sum}) and (\ref{stat-sum}) respectively.}
\label{fig2}
\end{figure*}

\begin{figure*}
\includegraphics[width=12.324cm] {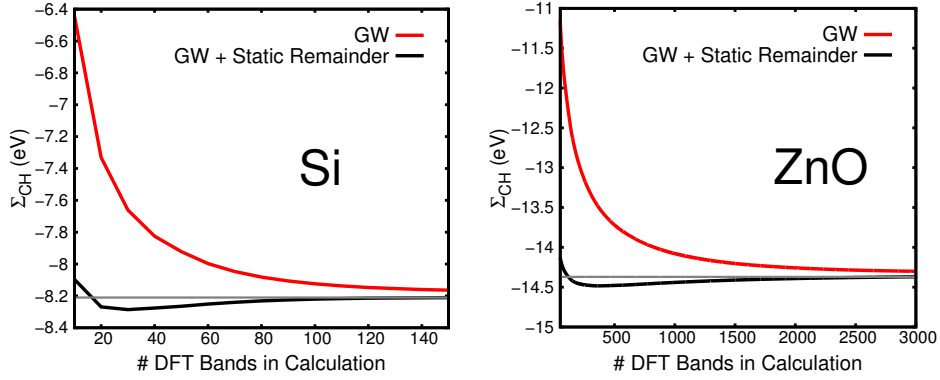}
\caption{Coulomb-hole energies of the valence band maximum in Si (left) and ZnO (right) in the modified static-remainder approach compared to the energies from the standard approach of truncating the Coulomb-hole summation in Eq. \ref{ch-sum} as a function of the number of DFT bands. In the static-remainder approach the summation is also truncated at the same number of bands but the modified static remainder is added to the sum. A 5x5x5 and 5x5x4 {\bf k}-point grid is used in Si and ZnO respectively. The grey lines represent the result using the maximum number of bands and the static remainder included.
}
\label{fig3}
\end{figure*}

A comparison between the convergence of the residual value of the GW expression (Eq. \ref{ch-sum}) and 1/2 of the static COHSEX approximation (Eq. \ref{stat-sum}) for the Coulomb-hole contribution to the electron self-energy starting at some $n_{\text{DFT}}$ is shown in Fig. \ref{fig2} for the valence band maximum (VBM) in Silicon and the highest occupied molecular orbital (HOMO) in the silane molecule. The figure shows the cumulative contributions of the high-energy orbitals to $\Sigma_{\text{CH}}$ for both the GW operator and the 1/2 static COHSEX operator for orbitals above 12 and 100 for silicon and silane, respectively.  The residual value of the 1/2 static COHSEX results reproduce the equivalent GW curves extremely well.  Therefore, replacing the GW operator with the modified static remainder in Eq. \ref{static-remainder} yields very good agreement with a fully converged GW calculation. This justifies the truncation of the partial sum in Eq. \ref{ch-sum} and the addition of the modified static remainder correction.  For both silicon and silane, one can get a converged $\Sigma_{\text{CH}}$ to within 100 meV with less than 10\% of the original number of empty orbitals required.  This is a very high level of accuracy considering the modified static correction in both cases is greater than 1 eV. Even higher accuracy may be reached if one increases the number of actual DFT empty orbitals used. In the case of Si, an accurate $\Sigma_{\text{CH}}$ (within 100 meV) can be reached with the use of only 10 empty bands when a 5x5x5 {\bf k}-point grid was used. Furthermore, the convergence for this approach is nearly monotonic in terms of the number of empty Kohn-Sham orbitals employed in the calculation. Figure \ref{fig3} shows the convergence of the modified static remainder corrected $\Sigma_{\text{CH}}$ and the uncorrected GW $\Sigma_{\text{CH}}$ as a function of DFT empty bands used for Si an ZnO - the grey line corresponds the "best guess" final value using the static-remainder on top of the largest number of real DFT bands.

In table I, we show the convergence behavior with respect to empty orbitals of our GW+static-remainder approach compared to a traditional GW approach for bulk Si, MgO, ZnO, and solid Ar. In all cases, the modified static-remainder approach significantly improves the convergence rates. An accuracy of less than 100 meV in the absolute energies can be typically reached with only a few conduction bands. 

\begin{table}
\begin{center}
\begin{tabular}{ | c | c | c | c | c | }  \hline 
Si - $\Gamma$ & 10 Bands & 40 Bands & 80 Bands & 160 Bands \\ \hline
n=1      & -5.99 & 6.30 & 6.47 & 6.53 \\
n=1 (SR) & -6.56 & 6.55 & 6.55 & 6.55 \\
n=4      & 6.19 & 5.62 & 5.30 & 5.15 \\
n=4 (SR) & 4.99 & 4.99 & 5.04 & 5.08 \\
n=5      & 9.46 & 8.90 & 8.60 & 8.48 \\
n=5 (SR) & 8.35 & 8.33 & 8.39 & 8.41 \\
n=10      & 15.23 & 14.48 & 14.10 & 13.94 \\
n=10 (SR) & 13.76 & 13.73 & 13.80 & 13.84 \\
\hline
\hline
ZnO - $\Gamma$ & 100 Bands & 500 Bands & 1500 Bands & 3000 Bands \\ \hline
n=26 & 6.11 & 4.48 & 4.00 & 3.90 \\
n=26 (SR) & 3.88 & 3.73 & 3.80 & 3.82 \\
n=27 & 7.97 & 7.32 & 7.22 & 7.21 \\
n=27 (SR) & 7.22 & 7.19 & 7.20 & 7.21 \\
\hline
\hline
MgO - $\Gamma$ & 50 Bands & 200 Bands & 450 Bands & 900 Bands \\ \hline
n=4 & -2.09 & -2.86 & -2.95 & -2.96 \\
n=4 (SR) & -3.15 & -3.02 & -2.97 & -2.97 \\
n=5 & 5.10 & 4.81 & 4.78 & 4.78 \\
n=5 (SR) & 4.70 & 4.77 & 4.78 & 4.78 \\
\hline
\hline
Ar - $\Gamma$ & 50 Bands & 150 Bands & 375 Bands & 750 Bands \\ \hline
n=4 & -7.64 & -8.19 & -8.39 & -8.42 \\
n=4 (SR) & -8.52 & -8.50 & -8.43 & -8.43 \\
n=5 & 5.26 & 5.38 & 5.411 & 5.42 \\
n=5 (SR) & 5.45 & 5.43 & 5.42 & 5.42 \\
\hline	
\end{tabular}
\end{center}
\caption{Convergence of the $E_{QP}$ (in eV) within the $G_0W_0$ approximation with respect to the number of DFT orbitals used in the Coulomb-hole summation for several material systems. Here n refers to the band index and (SR) refers to the addition of the static remainder.}
\label{timeBreakdown}
\end{table}

\begin{figure}
\includegraphics[width=6.324cm] {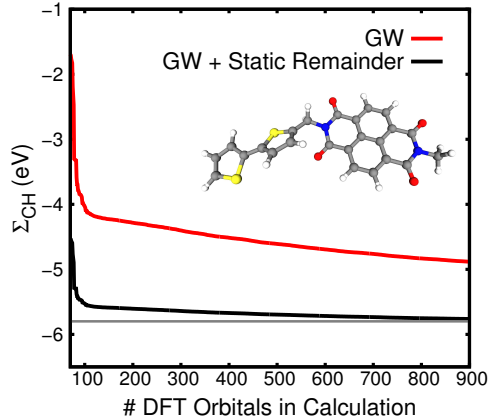}
\caption{Coulomb-hole part of the self-energy, with and without the static-remainder, for the highest occupied molecular orbital (HOMO) of the BND (bithiophene naphthalene diimide - shown in inset) molecule as a function of the number of DFT orbitals included in the Coulomb-hole sum. The grey line represents the result using the maximum number of bands and the static remainder included.}
\label{fig4}
\end{figure}

To test the modified static reminder approach on a large molecular system, we compute the Coulomb-hole contribution to the self-energy for the BND (bithiophene naphthalene diimide) molecule containing 46 atoms \cite{tao09}. The supercell was set to 76.93 x 36.31 x 20.18 atomic units. The calculations were done with a 60 Rydberg wavefunction cutoff and a 6 Rydberg dielectric matrix cutoff. The polarizability was computed with 953 orbitals (78 occupied + 875 empty orbitals up to 1 Rydberg cutoff in DFT eigenvalues), and the Coulomb-hole part of the self-energy was evaluated as a function of the number of orbitals, as shown in Fig. \ref{fig4}. One can see that the Coulomb-hole term computed with 953 orbitals without the addition of the remainder is only converged to within 1 eV. Including the static remainder correction improves the convergence to better than 0.1 eV.

In conclusion, we have presented a modified static remainder approach that reduces the number of empty states involved in evaluating $\Sigma_{\text{CH}}$ by over an order of magnitude.  This approach is particularly useful when applying the GW method to molecules and other nanostructures where absolute energies, as opposed to just energy gaps, are desired. A limitation of this method is that it does not address the problem of the sum over empty states required in evaluating the dielectric matrix (Eq. \ref{chi}). However, the dielectric matrix converges faster than the absolute energies of $\Sigma_{\text{CH}}$ for many solids \cite{shih10} and can more easily be replaced by calculation using the density functional perturbation theory approaches.  Our approach here shows nearly monotonic convergence towards the converged GW $\Sigma_{\text{CH}}$ values and can be implemented in a simple and automatic way in standard GW computer codes.

J.D. and M.J. acknowledge support from the Director, Office of Science, Office of Basic Energy Sciences, Materials Sciences and Engineering Division, U.S. Department of Energy under Contract No. DE-AC02-05CH11231. G.S. acknowledges support under National Science Foundation Grant No. DMR10-1006184. Computational resources have been provided by NSF through TeraGrid resources at NICS and by DOE at Lawrence Berkeley National Laboratory's NERSC facility.


\end{document}